\documentclass[review,11pt]{elsarticle}

\usepackage{lineno,hyperref}
\modulolinenumbers[5]

\journal{Journal of \LaTeX\ Templates}

\newcommand{\sgn}{{\mathop{\rm sgn}}}
\renewcommand{\vec}[1]{\mbox{\boldmath $#1$}}  

\newcommand{\mymat}[4]
       {\left(\!\!\begin{array}{cc}#1&#2\\#3&#4\end{array}\!\!\right)}
\newcommand{\mymatt}[9]
       {\left(\!\!\begin{array}{ccc}#1&#2&#3\\
                                    #4&#5&#6\\#7&#8&#9\end{array}\!\!\right)}
\newcommand{\rmD}{{\rm D}}

\newcommand{\rme}{{\rm e}}
\newcommand{\rmi}{{\rm i}}
\newcommand{\ep}{\epsilon}

\newcommand{\be}{\begin{equation}}
\newcommand{\ee}{\end{equation}}
\newcommand{\beq}{\begin{equation}}
\newcommand{\eeq}{\end{equation}}
\newcommand{\bea}{\begin{eqnarray}}
\newcommand{\eea}{\end{eqnarray}}
\newcommand{\bean}{\begin{eqnarray*}}
\newcommand{\eean}{\end{eqnarray*}}
\newcommand{\eqref}[1]{(\ref{#1})}

\newcommand{\calD}{{\mathcal D}}
\newcommand{\calM}{{\mathcal M}}
\newcommand{\calR}{{\mathcal R}}
\newcommand{\calS}{{\mathcal U}}
\newcommand{\calU}{{\mathcal X}}
\newcommand{\calV}{{\mathcal V}}

\usepackage{graphicx}
\usepackage[dvips]{epsfig}

\begin{document}

\begin{frontmatter}

\title{A third-order class-D amplifier with and
without ripple compensation}

\author[add1]{Stephen M. Cox}
\ead{stephen.cox@nottingham.ac.uk}
\address[add1]{School of Mathematical Sciences, University of Nottingham,
              University Park, Nottingham~NG7~2RD, United~Kingdom}

\author[add2]{H. du Toit Mouton}
\ead{dtmouton@sun.ac.za}
\address[add2]{Department of Electrical and Electronic Engineering,
              Stellenbosch University, Private~Bag~X1, Matieland~7602,
              South~Africa}

\begin{abstract}

We analyse the nonlinear behaviour of a third-order class-D amplifier, and 
demonstrate the remarkable effectiveness of the recently introduced ripple 
compensation (RC) technique in reducing the audio distortion of the device. The 
amplifier converts an input audio signal to a high-frequency train of 
rectangular pulses, whose widths are modulated according to the input signal 
(pulse-width modulation) and employs negative feedback.  After determining the 
steady-state operating point for constant input and calculating its stability, 
we derive a small-signal model (SSM), which yields in closed form the transfer 
function relating (infinitesimal) input and output disturbances. This SSM shows 
how the RC technique is able to linearise the small-signal response of the 
device.  We extend this SSM through a fully nonlinear perturbation calculation 
of the dynamics of the amplifier, based on the disparity in time scales between 
the pulse train and the audio signal. We obtain the nonlinear response of the 
amplifier to a general audio signal, avoiding the linearisation inherent in the 
SSM; we thereby more precisely quantify the reduction in distortion achieved 
through RC. Finally, simulations corroborate our theoretical predictions and 
illustrate the dramatic deterioration in performance that occurs when the 
amplifier is operated in an unstable regime. The perturbation calculation is 
rather general, and may be adapted to quantify the way in which other nonlinear 
negative-feedback pulse-modulated devices track a time-varying input signal 
that slowly modulates the system parameters.

\end{abstract}


\end{frontmatter}


\section{Introduction}

Class-D amplifiers are an important technological device and are widely 
used in mobile electronic devices, principally because of their 
exceptional efficiency~\cite{bd10}, which helps improve battery life. 
They operate by converting an audio signal to a high-frequency train of 
rectangular pulses whose widths are modulated in a manner that depends 
on the audio signal (pulse-width modulation, PWM)~\cite{B53}. They thus 
inherently involve dynamics on two different time scales and so are 
particularly amenable to analysis by perturbation methods.

To mitigate the influence of noise, designs for class-D amplifiers 
generally include some form of negative feedback, and hence may be 
modelled mathematically as piecewise-smooth dynamical 
systems~\cite{BBCK08,gc98}. However, while theoretical interest in such 
systems has primarily focused on the existence, stability and 
bifurcations of various steady-state operating points (see, for 
example,~\cite{BBCK08,gc98,amzg,azm,F97}), here we are principally concerned 
with the regime of most practical interest, which is the nonlinear 
response to a relatively slowly varying audio input. Our goal is to 
understand the way in which the pulse-modulated system tracks a slowly 
varying input signal, with particular focus on the low-frequency 
components of the amplifier output.

In recent publications, we have explored the operation of relatively 
simple first- and second-order designs of class-D amplifier (with, 
respectively, one or two integrators in the feedback 
path)~\cite{cc05,clt13,CM15,cty11,cygt} and we have illustrated the 
ripple compensation (RC) technique in the first-order case~\cite{CM15}, 
However, in all these cases, the operation of the feedback loop is 
simple enough that the entire mathematical model may be reduced to one 
or two nonlinear scalar difference equations for the switching times of 
the output pulse-train. Here we treat a higher-order design, for which 
reduction to such a simple system is no longer possible, and the problem 
is formulated instead as a nonlinear system of difference equations with 
slowly varying forcing. Specifically, here we treat a more realistic 
mathematical model, with a third-order compensator in the feedback loop, 
and a second-order output filter, so that the state-space model for the 
amplifier is five-dimensional.

A key diagnostic of practical interest is the audio distortion that 
arises due to the nonlinearity of the switching in the negative feedback 
loop. Over the years, many techniques have been devised by engineers to 
reduce this inherent distortion, and thereby to improve the fidelity of 
the audio reproduction. The present paper is dedicated to examining the 
theoretical basis of the ripple compensation (RC) technique, which 
appears to be a particularly effective means of eliminating significant 
elements of the distortion~\cite{MP09,P06}. Calculation of the audio 
distortion is achieved by first considering the dynamics of the 
amplifier, specifically determining the relationship between the audio 
input and the switching times of the output pulse-train, then from these 
switching times determining the audio content of the output. Despite the 
algebraically involved nature of the problem, we are able to give 
explicit formulas for the principal components of the audio output of 
the amplifier; these expressions make clear the contribution of the RC 
technique towards linearising the output.

In Section~\ref{sec:mathform}, we describe the amplifier treated in this 
paper and the RC technique; we also formulate the state-space model for 
the device. In Section~\ref{sec:stst}, we calculate the steady-state 
(time-periodic) operating point of the device in response to a constant 
input, and briefly consider its stability in Section~\ref{sec:stab}. 
This stability calculation informs the choice of parameter values for 
later simulations (our goal is not to explore instability and 
bifurcation, rather to ensure that the device is operated in a stable 
regime of practical relevance). In Section~\ref{sec:ss}, we develop a 
small-signal model which yields a transfer function relating small 
disturbances at the input to the consequential small disturbances at the 
output. This small-signal model allows us to deduce certain aspects of 
the behaviour of the device in response to a full audio signal, and in 
particular shows the linearising effect of RC. The principal results of 
this paper are contained in Section~\ref{sec:nl}, where we carry out a 
large-signal perturbation calculation of the amplifier output, where the 
perturbation parameter is proportional to the ratio between typical time 
scales for the output switching and the audio signal. We corroborate our 
theoretical results with corresponding simulations, which are presented 
in Section~\ref{sec:res}. Besides simulations in the stable regime of 
practical interest, we illustrate the calamitous sudden increase in the 
audio distortion that arises when the parameters of the device are 
poorly chosen, so that the steady-state operating point is unstable. 
Finally, we close, in Section~\ref{sec:conc}, by summarising our results 
and emphasising that our analysis --- in particular the asymptotic 
calculation of Section~\ref{sec:nl} --- may be applied to a wide range 
of negative-feedback pulse-modulated systems.

\clearpage
\section{Mathematical formulation}\label{sec:mathform}

\begin{figure}
\begin{center}
\epsfig{file=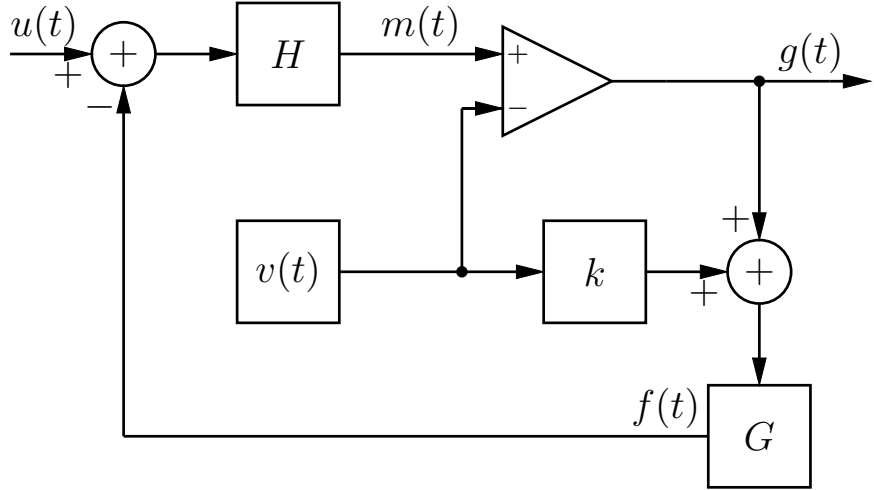,width=0.7\linewidth}
\end{center}

\caption{Third-order amplifier. The input audio signal is $u(t)$. $G$ 
represents the output filter. There is ripple compensation (RC) if $k=1$; 
otherwise, if $k=0$, there is no RC. The amplifier output is $g(t)$. The 
compensator is denoted by $H$. The output of the compensator, denoted by 
$m(t)$, is fed into the positive input of the comparator, whose negative 
input receives the sawtooth wave $v(t)$. The output takes the values 
$\pm1$ according to the sign of $m(t)-v(t)$.}

\label{fig:diag}
\end{figure}

\begin{figure}
\begin{center}
\epsfig{file=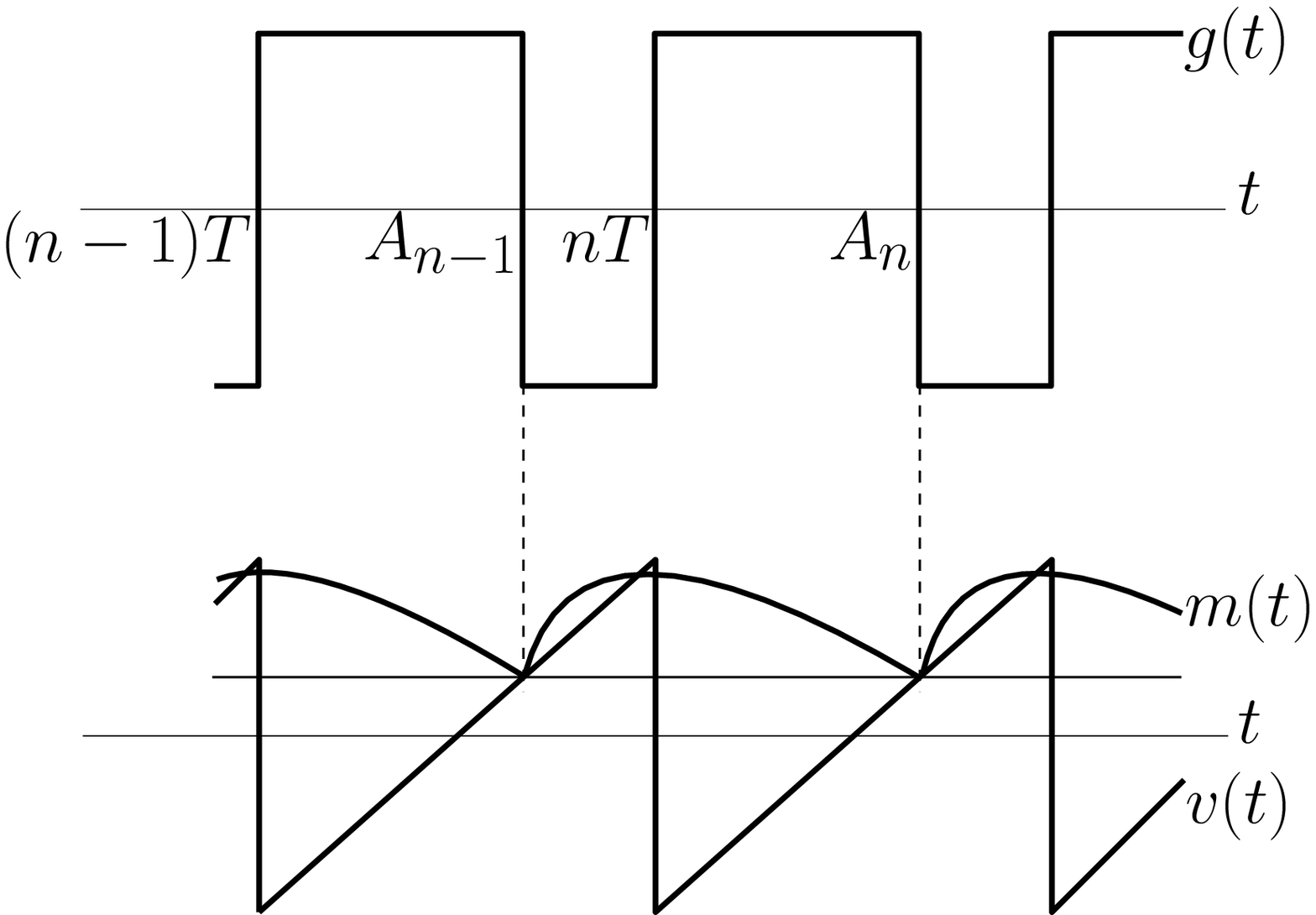,width=0.5\linewidth}

\vspace*{2ex}

\epsfig{file=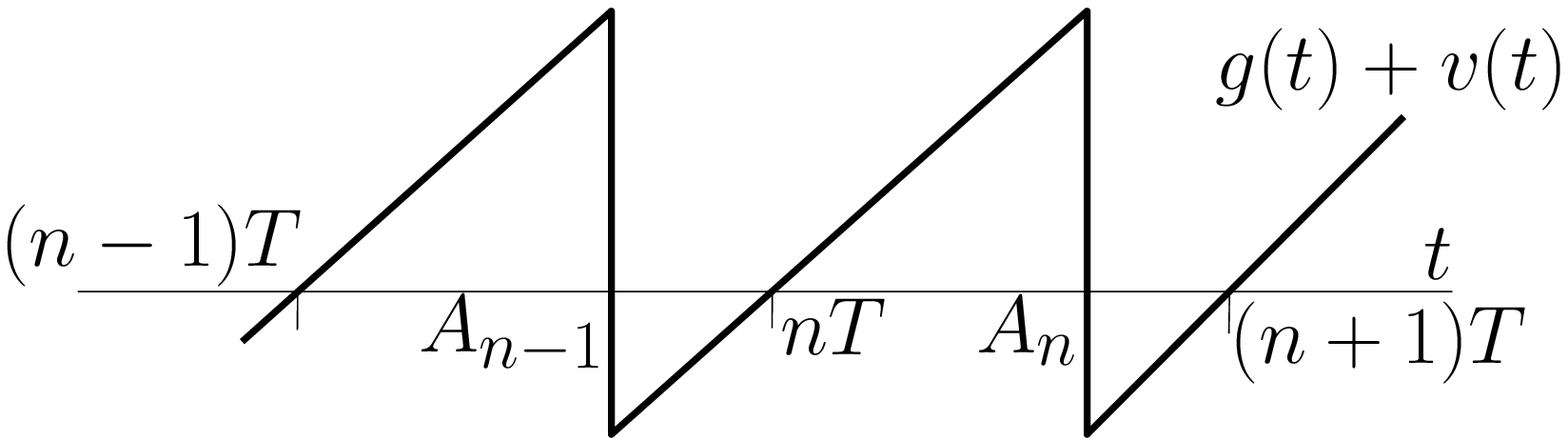,width=0.5\linewidth}
\end{center}

\caption{Signals $g(t)$, $v(t)$, $m(t)$ and $g(t)+v(t)$. The rising 
switching edges of $g(t)$ occur regularly, at times $t=nT$; the falling 
switching edges occur at times $t=A_n$, where $m(A_n)=v(A_n)$. (The 
signal $m(t)$ is for illustrative purposes.) The signal $g(t)+v(t)$ is a 
linear ramp, with falling switching edges at times $t=A_n$.}

\label{fig:gvk}
\end{figure}

Figure~\ref{fig:diag} shows the amplifier. The output $g(t)$ is a 
rectangular wave taking the values $\pm1$ according to
\[
g(t)=\sgn(m(t)-v(t)),
\]
where $m(t)$ and $v(t)$ are, respectively, the noninverting and inverting 
inputs of a comparator. The rising edges of $g(t)$ occur at regular 
intervals, where $t=nT$ and the constant $T$ is the period of the carrier 
wave $v(t)$. The falling edges of $g(t)$ occur at times that vary 
according to the output $m(t)$ of the compensator, as illustrated in
Figure~\ref{fig:gvk}. We denote these 
modulated down-switching times by $A_n$, so that
\beq
g(t)=\left\{\begin{array}{ll}+1&
            \mbox{for }      nT<t<A_n,\\
                             -1 &
            \mbox{for }      A_n<t<(n+1)T.
           \end{array}\right.
\label{eq:gA}
\eeq
The sawtooth carrier wave is given by 
\[
v(t)=-1+2(t-nT)/T
\]
for $nT\leq t<(n+1)T$ (with $v(t+T)=v(t)$ for all $t$),
and hence the condition for switching is
\beq
m(A_n)=-1+2a_n,\quad \mbox{where $a_n=(A_n-nT)/T$.}
\label{eq:switchm}
\eeq
The output filter $G$ receives as input $g(t)+kv(t)$, where $k$ is either 
$0$ or $1$. The choice $k=0$ indicates that RC is not applied; we note 
that in this case the filter input is $g(t)$, which is piecewise constant, 
switches up at times $t=nT$ and down at times $t=A_n$. Otherwise, the 
choice $k=1$ corresponds to the application of RC; the filter input is now 
$g(t)+v(t)$, which is a piecewise linear upwards ramp, which switches down 
at times $t=A_n$, as in Figure~\ref{fig:gvk}.

The key to the success of RC in reducing output distortion may be 
illustrated with the following observations regarding the steady-state 
$T$-periodic response of the amplifier to a constant input $u(t)=u_0$. 
When $k=0$, the steady-state duty cycle $a_n\equiv a$ varies according to 
the amplifier input $u_0$; for different values of $a$, the shape of 
$g(t)$ is thus different, and hence the shape of the compensator output 
varies accordingly. Put differently, the ripple depends on the input 
$u_0$. By contrast, with RC the shape of $g(t)+v(t)$ is same regardless of 
the value of $u_0$, except for a $u_0$-dependent time shift and the 
addition of a $u_0$-dependent constant to the values of $g(t)+v(t)$. Put 
differently, as we shall make precise below, in Section~\ref{sec:rip}, 
with RC the ripple is in essential respects independent of $u_0$.

\begin{figure}
\begin{center}
\epsfig{file=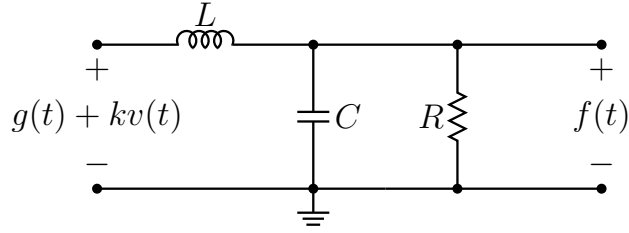,width=0.5\linewidth}
\end{center}
\caption{The second-order output filter, $G$.}
\label{fig:G}
\end{figure}

\begin{figure}
\begin{center}
\epsfig{file=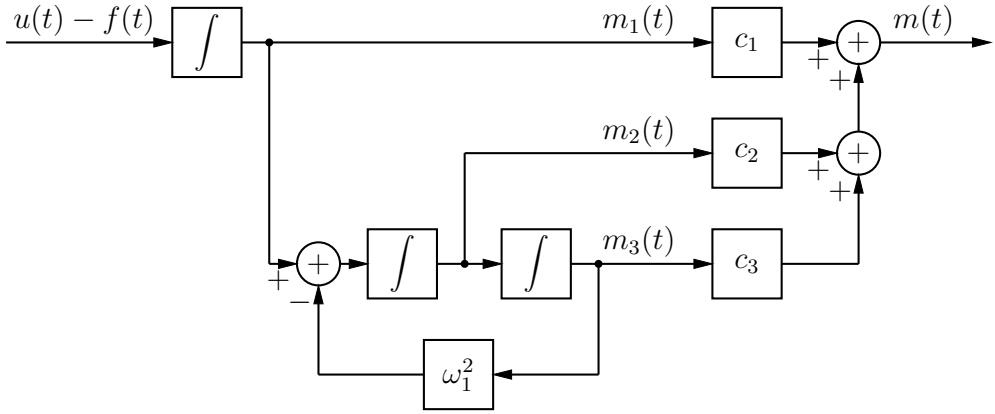,width=0.8\linewidth}
\end{center}
\caption{The third-order compensator, $H$.}
\label{fig:H}
\end{figure}

\subsection{State-space model}

We next present the ordinary differential equations that govern the 
operation of the device. We begin by examining the action of the 
low-pass filter $G$, shown in Figure~\ref{fig:G}. We let $\vec 
f(t)=(f(t),f'(t))^T$, where, as in the remainder of the paper, the 
superscript $T$ denotes the transpose of a vector or matrix. Then
 \beq
\vec f'(t)=M_2\vec f(t)+\frac{g(t)+kv(t)}{LC}(0,1)^T,
\label{eq:fode}
\eeq
where
\[
M_2=\mymat{0}{1}{-(LC)^{-1}}{-(RC)^{-1}}.
\]
The filter may equivalently be specified in terms of its (Laplace)
transfer function 
\[
G(s)=1/(LCs^2+Ls/R+1).
\]
Next we turn to the compensator $H$, which comprises a chain of integrators
with feed-forward summation and a local resonator feedback loop,
shown in Figure~\ref{fig:H}. We introduce 
the state vector $\vec m(t)=(m_1(t),m_2(t),m_3(t))^T$. Then
\beq
\vec m'(t)=M_3\vec m(t)+(u(t)-f(t))(1,0,0)^T,
\label{eq:mode}
\eeq
where
\[
M_3=\mymatt{0}{0}{0}{1}{0}{-\omega_1^2}{0}{1}{0}.
\]
The output of the compensator is
\[
m(t)=c_1m_1(t)+c_2m_2(t)+c_3m_3(t),
\]
for some constants $c_1$, $c_2$, $c_3$.
The (Laplace) transfer function of the compensator is
\[
H(s)=\frac{c_1}{sT}+\frac{c_2}{(\omega_1^2+s^2)T^2}+
\frac{c_3}{(\omega_1^2+s^2)sT^3}.
\]

We solve the systems \eqref{eq:fode} and \eqref{eq:mode} together, 
by introducing the state-space vector
\[
\vec x(t)=(m_1(t),m_2(t),m_3(t),f(t),f'(t))^T,
\]
which is governed by
\beq
\vec x'(t)=N\vec x(t)+u(t)\vec e_1+\frac{g(t)+kv(t)}{LC}\vec e_5,
\label{eq:uode}
\eeq
where  $\vec e_1=(1,0,0,0,0)^T$, \ldots, $\vec e_5=(0,0,0,0,1)^T$.
The matrix $N$ is partitioned as follows:
\[
N=\left(\begin{array}{cc}
M_3&
\begin{array}{cc}
-1&0\\0&0\\0&0
\end{array}
\\
\begin{array}{ccc}
0&0&0\\0&0&0
\end{array}
 & M_2
\end{array}
\right).
\]

To simplify the analysis here and in later sections, we next diagonalise 
$N$. We thus introduce the diagonal matrix $\Lambda$ of the eigenvalues 
of $N$, which are $0$, $\rmi\omega_1$, $-\rmi\omega_1$, 
$-\mu+\rmi\Omega$ and $-\mu-\rmi\Omega$, where
 \[
\mu=\frac{1}{2RC},\qquad\Omega=\sqrt{\frac{1}{LC}-\frac{1}{4R^2C^2}}.
\]
We also introduce a matrix $R$ whose columns are given, respectively,
by the corresponding right eigenvectors of $N$:
$\vec w_1$, \ldots, $\vec w_5$. Then
\beq
N=R\Lambda R^{-1}.
\label{eq:NR}
\eeq
The rows of $R^{-1}$ are the left eigenvectors of $N$:
$\vec v_1$, \ldots, $\vec v_5$.
Of particular utility in our analysis will be the left zero eigenvector
\beq
\vec v_1=(-(LC)^{-1},0,0,(RC)^{-1},1).
\label{eq:vN}
\eeq
Correspondingly, $\vec w_1=(-LC,0,-LC/\omega_1^2,0,0)^T$.

To simplify later notation when we integrate \eqref{eq:uode},
we introduce $\vec P_n(t)$ and $\vec Q_n(t)$, where
\beq
\vec P_0(t)=\rme^{Nt}\vec e_1,\quad\mbox{and}\quad
\vec P_{n+1}(t)=\int_0^t\vec P_n(\tau)\,{\rm d}\tau
\label{eq:Ps}
\eeq
for $n=0,1,\ldots$;
\[
\vec Q_0(t)=\rme^{Nt}\vec e_5,\quad\mbox{and}\quad
\vec Q_{n+1}(t)=\int_0^t\vec Q_n(\tau)\,{\rm d}\tau
\]
for $n=0,1,\ldots$.
We find
\[
\vec P_n(t)=\frac{t^n}{n!}\vec e_1+\phi_n(t)\vec e_2+\phi_{n+1}(t)\vec e_3,
\]
where $\phi_{-1}(t)=\cos\omega_1t$ and
\[
\phi_{n+1}(t)=\int_0^t\phi_n(\tau)\,{\rm d}\tau\quad
\mbox{for $n=-1,0,1,\ldots$.}
\]

We may readily integrate \eqref{eq:uode} over any time interval $[t_0,t_1]$,
to give
\beq
\vec x(t_1)=\rme^{N(t_1-t_0)}\vec x(t_0)+
\int_{t_0}^{t_1}\rme^{N(t_1-\tau)}\vec e_1u(\tau)\,{\rm d}\tau
+ \frac{1}{LC}\int_{t_0}^{t_1}\rme^{N(t_1-\tau)}\vec e_5
(g(\tau)+kv(\tau))\,{\rm d}\tau.
\label{eq:uint}
\eeq
For later purposes, our specific interest is in integrating 
\eqref{eq:uode} over the interval $[A_n,A_{n+1}]$ between successive 
falling edges of the amplifier output $g(t)$. To evaluate the first 
integral in \eqref{eq:uint}, we begin by writing
\[
\mathcal{I}_{u,n}\equiv
\int_{A_n}^{A_{n+1}}\rme^{N(A_{n+1}-\tau)}\vec e_1u(\tau)\,{\rm d}\tau.
\]
Then Taylor expansion of $u(\tau)$ about $\tau=A_n$ and repeated use
of integration by parts on the result, together with \eqref{eq:Ps}, gives
\beq
\mathcal{I}_{u,n}=\sum_{k=0}^\infty\vec P_{k+1}(A_{n+1}-A_n)u^{(k)}(A_n),
\label{eq:Isn}
\eeq
where the superscript denotes the $k$-th derivative. The second integral 
in \eqref{eq:uint} is similarly found (after just one integration by 
parts, to deal with $v(\tau)$) to be
\bean
\mathcal{I}_{g,v,n}&\equiv&\int_{A_n}^{A_{n+1}}\rme^{N(A_{n+1}-\tau)}\vec e_5
(g(\tau)+kv(\tau))\,{\rm d}\tau\\
&=&2(1-k)\vec Q_1(a_{n+1}T)+(-1-k+2ka_n)\vec Q_1(A_{n+1}-A_n)
+\frac{2k}{T}\vec Q_2(A_{n+1}-A_n).
\eean
Assembling these results, we thus arrive at the discrete-time model
\beq
\vec x(A_{n+1})=\rme^{N(A_{n+1}-A_n)}\vec x(A_n)+
\mathcal{I}_{u,n} +\frac{1}{LC}\mathcal{I}_{g,v,n},
\label{eq:u1}
\eeq
together with the switching condition \eqref{eq:switchm}, which
becomes
\beq
\vec \gamma^T\vec x(A_{n+1})=-1+2a_{n+1},
\label{eq:ctu}
\eeq
where
\[
\vec \gamma^T=\left(c_1,c_2,c_3,0,0\right).
\]

The system \eqref{eq:u1}, \eqref{eq:ctu} forms the basis for our 
mathematical analysis of the amplifier. We note that this system may be 
reduced to a single (fifth-order, nonlinear) scalar difference equation 
for $a_n$ (cf.~\cite{p04}). However, while we shall make use of a related 
reduction later, in Section~\ref{sec:nl}, the bulk of our analysis 
concerns the formulation in \eqref{eq:u1}, \eqref{eq:ctu}.

\section{Steady-state operation}\label{sec:stst}

We begin our analysis of the mathematical model set out above by 
examining its steady-state behaviour in response to a constant input. 
This is necessary in order for us to choose suitable parameter values 
for our simulations, where the steady-state response should be stable; 
it also sheds some light on the operation of RC.

We thus suppose that $u(t)=u_0$ and that all signals are $T$-periodic. In 
particular all duty cycles are equal, with $a_n\equiv a$.
In such steady-state operation, \eqref{eq:u1} becomes
\beq
(I_5-\rme^{NT})\vec x(aT)=\vec\Phi(a,T),
\label{eq:uss}
\eeq
where
\[
\vec\Phi(a,T)=u_0\vec P_1(T)
+\frac{1}{LC}\bigg\{2(1-k)\vec Q_1(aT)+(-1-k+2ka)\vec Q_1(T)
+\frac{2k}{T}\vec Q_2(T)\bigg\},
\]
and the switching condition is $\vec\gamma^T\vec x(aT)=-1+2a$.
Note that the matrix on the left-hand side of \eqref{eq:uss} is singular,
and after left-multiplying by $\vec v_1$, given in \eqref{eq:vN},
we obtain the solvability constraint
\beq
\vec v_1\vec\Phi(a,T)=0;
\label{eq:vPhi}
\eeq
this then yields the duty-cycle condition
\beq
a=\textstyle\frac12(1+u_0),
\label{eq:duty}
\eeq
which expresses the fact that the time-averaged output $\langle g(t)
\rangle=u_0$.

Once the duty-cycle condition \eqref{eq:vPhi} has been imposed, it 
remains to determine $\vec x(aT)$, from which the entire periodic 
solution may subsequently be obtained using \eqref{eq:uint}. This is 
accomplished by replacing one row (for example, the first) of the vector 
equation~\eqref{eq:uss} with the switching condition \eqref{eq:ctu}; 
thus we solve
\[
\tilde{M}\vec x(aT)=\tilde{\vec\Phi},
\]
where
\[
{\tilde M}_{ij}=\left\{\begin{array}{ll}c_j &\mbox{for $i=1$ and $j=1,2,3$,}\\
0 &\mbox{for $i=1$ and $j=4,5$,}\\
(I_5-\rme^{NT})_{ij} &\mbox{for $i=2,3,4,5$}
\end{array}\right.
\]
and
\[
\tilde{\Phi}_{i}=\left\{\begin{array}{ll}-1+2a &\mbox{for $i=1$,}\\
\Phi_i(a,T) &\mbox{for $i=2,3,4,5$.}
\end{array}\right.
\]
Explicit formulas for the steady-state solution are too algebraically involved
to record here.

A quantity of particular significance in our later stability calculation and
in our development of a model for small disturbances to the steady state is the
slope of the compensator output at the modulated switching instant; thus we
note from~\eqref{eq:uode} that
\beq
\vec\gamma^T\vec x'(aT)=\vec\gamma^TN\vec x(aT)+c_1u_0.
\label{eq:u'}
\eeq

\subsection{Relation between steady-state operating points for different 
inputs (with ripple compensation)} \label{sec:rip}

With RC, there is a particularly simple relationship between the 
steady-state operating points for different values of the input $u_0$, 
which we elucidate in this section. The simplicity of this relationship 
underpins the effectiveness of RC in reducing the amplifier's inherent 
distortion.

We consider two different $T$-periodic steady-state solutions, with different
values of $u_0$, and hence different switching times for $g(t)+v(t)$.
For each solution, the duty-cycle condition \eqref{eq:duty} is satisfied,
together with the state-space equation \eqref{eq:uode} and the switching
condition \eqref{eq:ctu}. Thus the two solutions, $\vec x_a(t)$ and
$\vec x_b(t)$, satisfy
\beq
\begin{array}{l}
\displaystyle{\vec x}'_a(t)
=N\vec x_a(t)+(2a-1)\vec e_1+\frac2{LC}(-1+t/T)\vec e_5
\qquad\mbox{for $aT\leq t<aT+T$},\\
\vec\gamma^T\vec x_a(aT)=-1+2a
\end{array}
\eeq
and
\beq
\begin{array}{l}
\displaystyle{\vec x}'_b(t)
=N\vec x_b(t)+(2b-1)\vec e_1+\frac2{LC}(-1+t/T)\vec e_5
\qquad\mbox{for $bT\leq t<bT+T$},\\
\vec\gamma^T\vec x_b(bT)=-1+2b.
\end{array}
\eeq
To demonstrate the relationship between the two solutions, we introduce
the $T$-periodic quantity
\[
\vec\Delta(t)=\vec x_b(t+(b-a)T)-\vec x_a(t),
\]
which satisfies
\beq
\begin{array}{l}
\displaystyle{\vec\Delta}'(t)=N\vec\Delta(t)+
2(b-a)\left(\vec e_1+\frac1{LC}\vec e_5\right)
\qquad\mbox{for $aT\leq t<aT+T$},\\
\vec\gamma^T\vec\Delta(aT)=2(b-a).
\end{array}
\label{eq:19}
\eeq
The general solution to the ODE in \eqref{eq:19} is
\beq
\vec \Delta(t)=\rme^{N(t-aT)}\vec\Delta(aT)+2(b-a)
\int_{aT}^{t}\rme^{N(t-\tau)}\vec \epsilon\,{\rm d}\tau,
\label{eq:Dint}
\eeq
where
\[
\vec\epsilon=\vec e_1+\frac{1}{LC}\vec e_5.
\]
From \eqref{eq:Dint}, we see that
\[
N\vec\Delta(t)=\rme^{N(t-aT)}N\vec\Delta(aT) +2(b-a)
\left(\rme^{N(t-aT)}-I_5\right)\vec\epsilon,
\]
which may be rearranged as
\beq
\vec X(t)=\rme^{N(t-aT)}\vec X(aT),
\label{eq:XeX}
\eeq
where
\beq
\vec X(t)=N\vec\Delta(t)+2(b-a)\vec\epsilon.
\label{eq:X}
\eeq
Since $\vec X(t)$ is $T$-periodic, it follows from \eqref{eq:XeX} with
$t=aT+T$ that
\beq
\left(\rme^{NT}-I_5\right)\vec X(aT)=\vec 0.
\label{eq:eIX}
\eeq
Hence $\vec X(aT)=X_3\vec w_1$ for some constant $X_3$.
The value of $X_3$ may be determined from \eqref{eq:X}: we see that
\[
\vec v_1\vec X(aT)=\vec v_1 N\vec\Delta(aT)+2(b-a)\vec v_1\vec\epsilon=0,
\]
hence (considering leftmost and rightmost sides of this equation)
$X_3=0$. Thus $\vec X(aT)=\vec 0$, and from~\eqref{eq:XeX} it
follows that $\vec X(t)\equiv\vec 0$, so that,
from \eqref{eq:X},
\[
\vec\Delta(t)=(\omega_1^2,0,1,0,0)^T\Delta_3(t)+(0,0,0,2(b-a),0)^T,
\]
for some $\Delta_3(t)$. From
the switching condition in \eqref{eq:19}, we see that $(c_1\omega_1^2+c_3)
\Delta_3(t)=2(b-a)$. Hence $\vec\Delta(t)$ is in fact constant, with
\[
\vec\Delta=\frac{2(b-a)}{c_1\omega_1^2+c_3}\left(
\omega_1^2,
0,
1,
c_1\omega_1^2+c_3,
0
\right)^T.
\]
In summary, we have shown that, for $aT\leq t<aT+T$,
\[
\vec x_b(t+(b-a)T)=\vec x_a(t)+\vec\Delta.
\]

Since the two steady-state solutions differ by the addition of a constant 
vector, and by time-shifting, the derivatives of each solution around the 
modulated switching instant agree: more specifically,
\[
\vec\gamma^T\vec x_b'(bT)=\vec\gamma^T\vec x_a'(aT).
\]
This fact has significant consequences, as we shall see later, in 
Section~\ref{sec:ss}, when we show how it leads to a linearisation 
of the small-signal model.

\section{Stability of the steady-state operating point}
\label{sec:stab}

Our interest is in stable operation of the amplifier, so we provide 
just a brief discussion of stability considerations. Following Aizerman 
and Gantmakher~\cite{ag58} (see also, for 
example,~\cite{BBCK08,F97,FA02}), we suppose that the input $u(t)=u_0$ 
is fixed, and consider the growth or decay of a perturbation to the 
steady state over the interval $t\in[0,T]$.  We write
\[
\vec x(t)=\bar{\vec x}(t)+\Delta\vec x(t),\qquad
a_0=a+\Delta a,
\]
where $\bar{\vec x}(t)$ is the steady-state solution with duty cycle $a$.
Then, upon linearising in small disturbances, we find
\beq
\Delta \vec x(T)=
\underbrace{\rme^{N(1-a)T}
\left(I_5+\frac{T\kappa}{LC}\vec e_5\vec\gamma^T\right)
\rme^{NaT}}_{\equiv\mathcal{M}}
\Delta \vec x(0),
\label{eq:M}
\eeq
where
\beq
\kappa=(1-\textstyle\frac12T\vec \gamma^T\bar{\vec x}'(aT))^{-1},
\label{eq:Dau}
\eeq
and the quantity $\vec\gamma^T\bar{\vec x}'(aT)$ may be obtained
from~\eqref{eq:u'}.
The stability of the steady-state operating point is thus determined by 
the eigenvalues of the matrix $\mathcal{M}$~\cite{BBCK08}. Note that the 
sole difference between the RC and no-RC versions of $\calM$ lies in the
value of $\kappa$.

The eigenvalues $\mu$ of $\mathcal{M}$ satisfy
\beq
\det(\mathcal{M}-\mu I)=0.
\label{eq:det}
\eeq
We may derive an alternative equation for these eigenvalues
(cf.~\cite{F97,FA99,FA01}), which may be more useful in some cases,
by use of Sylvester's Determinant Theorem~\cite{h97}, which states that
$\det(I_n-AB)=\det(I_p-BA)$, where
$A$ is any $n\times p$ matrix and $B$ is any $p\times n$
matrix, and 
$I_n$ and $I_p$ are, respectively, $n\times n$ and $p\times p$
identity matrices.
We suppose, as is readily verified, that none of the eigenvalues of
$\exp(NT)$ are also eigenvalues of $\mathcal{M}$. Then
\[
\det(\mathcal{M}-\mu I)=\det(\rme^{NT}-\mu I+\vec\alpha\vec\beta^T)
=
\det(\rme^{NT}-\mu I)
\det(I+(\rme^{NT}-\mu I)^{-1}\vec\alpha\vec\beta^T),
\]
where
\[
\vec\alpha=\frac{T\kappa}{LC}\rme^{N(1-a)T}\vec e_5,\quad
\vec\beta^T=\vec\gamma^T\rme^{NaT}.
\]
Hence the eigenvalues $\mu$ of $\mathcal{M}$ satisfy
\beq
\det(I+(\rme^{NT}-\mu I)^{-1}\vec\alpha\vec\beta^T)=0
\label{eq:syl1}
\eeq
and so, by Sylvester's Determinant Theorem, they also satisfy
the equation
\[
1+\frac{T\kappa}{LC}\vec\gamma^T\rme^{NaT}(\rme^{NT}-\mu I)^{-1}
\rme^{N(1-a)T}\vec e_5=0,
\]
or, equivalently, 
\beq
1+\frac{T\kappa}{LC}\vec\gamma^TR
\mathop{{\rm diag}}(\rme^{\lambda_jT}/(\rme^{\lambda_jT}-\mu))R^{-1}\vec e_5=0,
\label{eq:detsca}
\eeq
where $\lambda_j$ are the eigenvalues of $N$. We use either 
\eqref{eq:det} or \eqref{eq:detsca} to choose parameter values for our 
simulations (see Section~\ref{sec:res}) so that the steady-state 
operating point is stable, unless otherwise stated.

\section{Small-signal model}\label{sec:ss}

The stability analysis of Section~\ref{sec:stab} may be generalised to 
give a small-signal model that relates small disturbances at the input to 
the corresponding small disturbances to the output. We suppose that some 
small time-dependent perturbation is superposed on an otherwise steady 
input, so that $u(t)=u_0+\Delta u(t)$. We introduce the notation $\bar 
A_n=(n+a)T$ for the unperturbed switching times, and write the perturbed 
switching times as $A_n=\bar A_n+\Delta a_n T$. We write $\vec x(t)= 
\bar{\vec x}(t)+\Delta \vec x(t)$ accordingly. We linearise in all small 
quantities.

In what follows, we assume that the $n$-th switching instant is 
delayed, so that $\Delta a_n>0$. This is simply to fix the time-ordering 
of various events; the resulting expressions for the small-signal model do 
not rely on this assumption.

We note that $\bar{\vec x}(t)$ depends on the value of $k$. By contrast,
for the perturbations, regardless of whether $k=0$ or $1$, we have the
following governing equations: on $(\bar A_n,\bar A_n+\Delta a_nT)$,
\[
\Delta{\vec x}'(t)=N\Delta\vec x(t)+\Delta u(t)\vec e_1+
\frac{2}{LC}\vec e_5;
\]
on $(\bar A_n+\Delta a_nT,\bar A_{n+1})$,
\[
\Delta{\vec x}'(t)=N\Delta\vec x(t)+\Delta u(t)\vec e_1.
\]
Integration of these differential equations in turn gives
\beq
\Delta\vec x(\bar A_{n+1})=
\rme^{NT}\left(\Delta\vec x(\bar A_n)+\frac{2\Delta a_nT}{LC}\vec e_5\right)
+\int_0^T\rme^{N\tau}\Delta u(\bar A_{n+1}-\tau)\vec e_1\,{\rm d}\tau.
\label{eq:ssu}
\eeq
The linearised switching condition \eqref{eq:ctu} yields
\beq
\Delta a_n={\textstyle\frac12}\kappa\vec\gamma^T\Delta \vec x(\bar A_n),
\label{eq:Da}
\eeq
where again $\kappa$ is given by \eqref{eq:Dau}.
We note that when $k=0$, $\kappa$ depends on $u_0$, whereas when $k=1$,
$\kappa$ is independent of $u_0$.

Through repeated integration by parts, as in the derivation of \eqref{eq:Isn},
it may be established, using \eqref{eq:ssu} and \eqref{eq:Da},
that
\beq
\Delta\vec x(\bar A_{n+1})=
{\mathcal N}\Delta\vec x(\bar A_n)
+\sum_{m=0}^\infty \vec P_{m+1}(T)\Delta u^{(m)}(\bar A_n),\label{eq:ssuP}
\eeq
where
\[
{\mathcal N}=
\rme^{NT}\left(I_5+\frac{T\kappa}{LC}\vec e_5\vec \gamma^T \right).
\]
(Note that ${\mathcal M}=\rme^{-NaT}{\mathcal N}\rme^{NaT}$, where
$\mathcal M$ is defined in \eqref{eq:M}, hence $\mathcal M$ and $\mathcal N$
are similar matrices and so share the same eigenvalues.)

A recurrence relation for the switching-time perturbation may now be
derived by premultiplying \eqref{eq:ssuP} by $\frac12\kappa\vec\gamma^T$
then using \eqref{eq:Da}.  A convenient formulation
for the solution may be obtained by introducing the derivative operator
$\rmD\equiv{\rm d}/{\rm d}t$ and using the Taylor expansion~\cite{B72,M33}
\[
\Delta\vec x(\bar A_{n+1})=\rme^{T\rmD}\Delta\vec x(\bar A_n),
\]
to give the formal solution
\beq
\Delta a_n=
{\textstyle\frac12}\kappa\vec\gamma^T
\left(\rme^{T\rmD}I_5-{\mathcal N}\right)^{-1}
\sum_{m=0}^\infty \vec P_{m+1}(T)\Delta u^{(m)}(\bar A_n).
\label{eq:Dan}
\eeq

The next step is to characterise the corresponding spectral components
of the output pulse-train. To this end, we let $x(t)$ be such that
$x(\bar A_n)=\Delta a_n$.  In view of \eqref{eq:Dan}, one particular choice
of $x(t)$ satisfies
\beq
x(t)={\textstyle\frac12}\kappa\vec\gamma^T
\left(\rme^{T\rmD}I_5-{\mathcal N}\right)^{-1}
\sum_{m=0}^\infty \vec P_{m+1}(T)\Delta u^{(m)}(t).
\label{eq:x}
\eeq

The final step in our derivation of
the small-signal model uses $x$ to reconstruct
the amplifier output.
From \eqref{eq:gA}, it follows that the Fourier transform of the full
output $g(t)$ is, for $\omega\neq0$,
\[
\hat g(\omega)=\int_{-\infty}^\infty\rme^{-\rmi\omega t}g(t)\,{\rm d}t
=\frac{2}{\rmi\omega}\sum_{n=-\infty}^\infty
\left(
\rme^{-\rmi\omega nT}-\rme^{-\rmi\omega A_n}
\right).
\]
By considering the difference between the
Fourier transform of the output with and without
perturbation, we find that the perturbation to the output has Fourier
transform
\beq
\Delta\hat g(\omega)=\frac{2}{\rmi\omega}\sum_{n=-\infty}^\infty
\left(
\rme^{-\rmi\omega \bar A_n}-\rme^{-\rmi\omega A_n}
\right).
\label{eq:Dghat}
\eeq
Linearisation in small perturbations then gives
\[
\rme^{-\rmi\omega A_n}=\rme^{-\rmi\omega \bar A_n}
\rme^{-\rmi\omega \Delta a_nT}\sim \rme^{-\rmi\omega \bar A_n}
(1-\rmi\omega \Delta a_nT).
\]
Thus \eqref{eq:Dghat} becomes
\beq
\Delta\hat g(\omega)=2T\sum_{n=-\infty}^\infty
\rme^{-\rmi\omega \bar A_n}x(\bar A_n)
=2\sum_{n=-\infty}^\infty
\rme^{-2\pi n\rmi a}\hat x(\omega-2\pi n/T),
\label{eq:gx}
\eeq
where the second equality follows from Poisson resummation~\cite{mf53}.

In particular, if we make the physically reasonable assumption that
the input perturbation $\Delta u(t)$ contains only audio frequencies, so
that $\Delta u(t)$ (and hence also $x(t)$) is band-limited, with
$\Delta\hat u(\omega)=\hat x(\omega)=0$ for $|\omega|\geq\pi/T$, 
then, from \eqref{eq:x} and \eqref{eq:gx},
\[
\Delta\hat g(\omega)=
\kappa\vec\gamma^T
\left(\rme^{\rmi\omega T}I_5-{\mathcal N}\right)^{-1}
\sum_{m=0}^\infty \vec P_{m+1}(T)(\rmi\omega)^m\Delta \hat u(\omega)
\]
for $|\omega|<\pi/T$.
To simplify the sum in this expression, we let
\[
\vec\sigma(T;\rmi\omega)=\sum_{m=0}^\infty \vec P_{m+1}(T)(\rmi\omega)^m,
\]
then note that in consequence
\[
\frac{{\rm d}\vec\sigma(T;\rmi\omega)}{{\rm d}T}
-\rmi\omega\vec\sigma(T;\rmi\omega)=\vec P_0(T),\qquad
\vec\sigma(0;\rmi\omega)=\vec 0.
\]
Solving this ODE, we thus have
\[
\vec\sigma(T;\rmi\omega)
=\sigma_1\vec e_1+\sigma_2\vec e_2+\sigma_3\vec e_3,
\]
where
\bean
\sigma_1&=&\frac{\rme^{\rmi\omega T}-1}{\rmi\omega},\\
\sigma_2&=&\frac{\rmi\omega\sin\omega_1T-\rmi\omega_1\sin\omega T
+\omega_1(\cos\omega_1T-\cos\omega T)}
{\omega_1(\omega^2-\omega_1^2)},\\
\sigma_3&=&\frac{\omega\sin\omega_1T-\omega_1\sin\omega T}
{\omega\omega_1(\omega^2-\omega_1^2)}+\rmi
\frac{\omega_1^2\cos\omega T-\omega^2\cos\omega_1T
+\omega^2-\omega_1^2}
{\omega\omega_1^2(\omega^2-\omega_1^2)}.
\eean
This, finally, yields the (input--output) transfer function,
from $\Delta \hat u(\omega)$ to $\Delta \hat g(\omega)$, which is
\beq
\kappa\vec\gamma^T
\left(\rme^{\rmi\omega T}I_5-{\mathcal N}\right)^{-1}
\vec\sigma(T;\rmi\omega),
\label{eq:tf}
\eeq
for ``audio frequencies'' (those less than $\pi/T$ in magnitude).

Without RC, this transfer function depends on $u_0$, through the value 
of $\kappa$ (both explicitly in \eqref{eq:tf} and implicitly through 
${\mathcal N}$) and so the small-signal model predicts an inherently 
nonlinear response for the amplifier. This is undesirable, since such 
nonlinearity leads to unwanted total harmonic distortion (THD) and 
intermodulation distortion (IMD)~\cite{cygt,ytcg}.

With RC, the transfer function is independent of $u_0$, so we expect it to 
provide an accurate characterisation of the input--output relation {\em 
even for inputs that are not small perturbations to some constant input}. 
This is a striking result, because it predicts an essentially linear 
behaviour for the amplifier.  More specifically, for an audio input 
$u(t)$, the small-signal model predicts an output with audio-frequency 
Fourier components given by
 \[ 
\hat 
g_a(\omega)=\kappa\vec\gamma^T \left(\rme^{\rmi\omega T}I_5-{\mathcal 
N}\right)^{-1} \vec\sigma(T;\rmi\omega)\hat u(\omega).
 \] 
In fact, as we shall demonstrate in the next section, the full audio 
output is {\em not quite} linearly related to the input: harmonics {\em 
are} generated, but from terms neglected in the small-signal 
linearisation (cf.~\cite{CM15}). An example of such a term is 
$((u')^2)'$, which involves a product of input derivatives; the 
contribution of such terms is, however, small~\cite{CM15}.

We next turn to a full calculation of the output that is not constrained 
by the linearisation inherent in the small-signal model.

\section{Fully nonlinear model}\label{sec:nl}

Our final calculation gives the nonlinear audio output in response to a 
general audio input. This calculation tracks the slowly changing 
operating point of the amplifier in response to its input, in sufficient 
detail to allow us to find the principal contributions to the output 
distortion. Of necessity, it avoids the traditional 
\textit{quasi-steady} engineering approximation, that the input to the 
amplifier is assumed constant over any switching cycle. We follow the 
structure of the calculation described in 
\cite{cc05,clt13,CM15,cty11,cygt}, although here the details are 
considerably more algebraically involved than in any of those previous 
cases. We emphasise that our approach may be readily adapted to other 
pulse-modulated feedback systems~\cite{gc98} with a slowly varying input 
parameters.

We apply a perturbation method based on the small parameter
\[
\ep=\omega T\ll1,
\]
where $\omega$ is a typical audio frequency. We introduce a correspondingly
scaled time
\[
\tau=\omega t=\ep t/T.
\]
Thus variations to the audio input occur on a time scale $\tau=O(1)$,
while the switching time scale has $\tau=O(\ep)$.
We introduce
\[
\calS(\tau)=u(t),
\]
so that $u^{(m)}(t)=(\ep/T)^m\calS^{(m)}(\tau)$. Our interest is in 
determining how solutions to the system \eqref{eq:u1}, \eqref{eq:ctu} 
track the slow parametric variation afforded by the input audio signal.

We first determine the way in which the switching times
depend on the audio input, then calculate the corresponding audio output.
To this end, we introduce functions $a$ and $\vec\calU$ such that
\[
a(\ep n)=a_n,\qquad
\vec \calU(\ep n)=\vec x(A_n)=\vec x((n+a(\ep n))T).
\]
Writing the difference equation \eqref{eq:u1} and switching condition 
\eqref{eq:ctu} in this notation, we find that each equation involves 
$a(\ep n)$ or $a(\ep(n+1))$. Clearly these equations are expected to hold 
only for integer values of $n$. However, as a mathematical device to 
enable a solution to be obtained, we seek to impose each equation for all 
real values of $n$ (since if we are able to do so then the equations 
certainly hold when restricted to integer $n$). Thus we set $\tau=\ep n$ 
and solve for all $\tau$ the following:
\beq
\vec \calU(\tau+\ep)=\rme^{Nd_\tau}\vec \calU(\tau)+\vec\Theta(\tau),\qquad
\vec\gamma^T\vec\calU(\tau)=-1+2a(\tau),
\label{eq:full}
\eeq
where
\bean
\vec\Theta(\tau)&=&
\sum_{m=0}^\infty \frac{\ep^m}{T^m}\vec P_{m+1}(d_\tau)
\calS^{(m)}(\tau+\ep a(\tau))+
\frac{2(1-k)\vec Q_1(a(\tau+\ep)T)}{LC}\\
&&{}-\frac{(1+k-2ka(\tau))\vec Q_1(d_\tau)}{LC}
+\frac{2k}{TLC}\vec Q_2(d_\tau)
\eean
and where
\[
d_\tau=(1+a(\tau+\ep)-a(\tau))T.
\]
The functions $a$ and $\vec\calU$ are then expanded in powers of $\ep$,
and coefficients of successive powers of $\ep$ equated in \eqref{eq:full}.

Given the algebraic complexity of the perturbation problem, it is useful
to reduce the problem from the six scalar equations represented in
\eqref{eq:full} to a single scalar equation,
for $a(\tau)$. To do so, we introduce
\beq
\vec\calV(\tau)=\rme^{-a(\tau)NT}\vec\calU(\tau).
\label{eq:VU}
\eeq
Then the first of \eqref{eq:full} becomes
\[
\vec\calV(\tau+\ep)=\rme^{NT}\vec\calV(\tau)+\rme^{-a(\tau+\ep)NT}
\vec\Theta(\tau),
\]
so that
\beq
\left(\rme^{\ep \rmD}I_5-\rme^{NT}\right)\vec\calV(\tau)=
\rme^{-a(\tau+\ep)NT}\vec\Theta(\tau),
\label{eq:VT}
\eeq
where throughout this section $\rmD$ denotes ${\rm d}/{\rm d}\tau$.

Then, using \eqref{eq:NR}, \eqref{eq:VT} may be written as
\[
R\left(\rme^{\ep \rmD}I_5-\rme^{\Lambda T}\right)R^{-1}\vec\calV(\tau)=
R\rme^{-a(\tau+\ep)\Lambda T}R^{-1}\vec\Theta(\tau),
\]
so that
\beq
\vec\calV(\tau)=
R\left(\rme^{\ep \rmD}I_5-\rme^{\Lambda T}\right)^{-1}
\left\{\rme^{-a(\tau+\ep)\Lambda T}R^{-1}\vec\Theta(\tau)\right\}.
\label{eq:Vis}
\eeq
Using \eqref{eq:VU} and \eqref{eq:Vis}, we see that the switching condition
in \eqref{eq:full} becomes
\beq
\vec\gamma^T\rme^{a(\tau)NT}R
\left(\rme^{\ep \rmD}I_5-\rme^{\Lambda T}\right)^{-1}
\left(\rme^{-a(\tau+\ep)\Lambda T}R^{-1}\vec\Theta(\tau)\right)=-1+2a(\tau),
\label{eq:a.}
\eeq
which is the promised single scalar equation for $a(\tau)$.

Several of the terms in this equation may readily be simplified, by 
introducing the Bernoulli numbers $B_n$ and the Bernoulli--Apostol 
functions $\beta_n$~\cite{a51}, which satisfy the following generating 
functions (for $\gamma\neq1$):
 \[
\frac{z}{\rme^z-1}=\sum_{n=0}^\infty\frac{B_n}{n!}z^n,\quad
\frac{z}{\gamma\rme^z-1}=\sum_{n=1}^\infty\frac{\beta_n(\gamma)}{n!}z^n.
 \]
Thus
\[
\calD\equiv\left(\rme^{\ep \rmD}I_5-\rme^{\Lambda T}\right)^{-1}=
\mathop{{\rm diag}}(\zeta_1,\zeta(\rmi\omega_1),\zeta(-\rmi\omega_1),
\zeta(-\mu+\rmi\Omega),\zeta(-\mu-\rmi\Omega)),
\]
where
\[
\zeta_1=\frac{1}{\ep \rmD}\sum_{n=0}^\infty\frac{B_n}{n!}(\ep \rmD)^n,\qquad
\zeta(z)=\rme^{-zT}
\sum_{n=0}^\infty\frac{\beta_{n+1}(\rme^{-zT})}{(n+1)!}(\ep \rmD)^n,
\]
and the equation for $a(\tau)$ simplifies from \eqref{eq:a.} to
\beq
\vec\gamma^TR\rme^{a(\tau)\Lambda T}
\calD
\left\{\rme^{-a(\tau+\ep)\Lambda T}R^{-1}\vec\Theta(\tau)\right\}=-1+2a(\tau).
\label{eq:ae}
\eeq

From the Fourier transform of \eqref{eq:gA}, it may be
deduced~\cite{cc05,clt13,CM15,cty11} that the audio contribution to the 
output (i.e., the contribution involving frequencies less than $\pi/T$) is
\beq
g_a(t)=-1+2
\sum_{n=0}^\infty\frac{(-\ep)^n}{(n+1)!}
\frac{{\rm d}^n a^{n+1}(\tau)}{{\rm d}\tau^n},
\label{eq:ga}
\eeq
and so, in principle, the required calculation is now clear:
we expand $a(\tau)$ in powers of $\ep$, as
\beq
a(\tau)=a_0(\tau)+\ep a_1(\tau)+O(\ep^2),
\label{eq:aexp}
\eeq
then solve \eqref{eq:ae} at successive powers of $\ep$ to find in turn the 
$a_n(\tau)$, finally substituting these expressions in \eqref{eq:ga} to 
determine the output. In practice, of course, the details are extremely 
algebraically cumbersome. The next section describes this calculation at 
the first two orders in $\epsilon$; these provide the principal 
contributions to the audio output.

\subsection{Calculation of the output to $O(\epsilon)$}
\label{sec:6.1}

From \eqref{eq:ga}, we see that the output takes the form
\beq
g_a(t)=-1+2a_0(\tau)+\ep g_1+O(\ep^2),
\label{eq:gais}
\eeq
where
\beq
g_1=2a_1(\tau)-2a_0(\tau)a_0'(\tau).
\label{eq:g1def}
\eeq
In solving \eqref{eq:ae} for $a_0(\tau)$ and $a_1(\tau)$, we need the following
expansions:
\bean
\rme^{a(\tau)\Lambda T}&=&\rme^{a_0(\tau)\Lambda T}+
\ep a_1(\tau)\rme^{a_0(\tau)\Lambda T}\Lambda T
+O(\ep^2),\\
\rme^{-a(\tau+\ep)\Lambda T}&=&\rme^{-a_0(\tau)\Lambda T}
-\ep(a_1(\tau)+a_0'(\tau))\rme^{-a_0(\tau)\Lambda T}\Lambda T+O(\ep^2).
\eean
We also expand
\[
\left(\rme^{\ep \rmD}I_5-\rme^{\Lambda T}\right)^{-1}=
\frac{1}{\ep \rmD}\Upsilon_{-1}+\Upsilon_0+O(\ep),
\]
where $\Upsilon_{-1}=\mathop{{\rm diag}}(1,0,0,0,0)$ and
\[
\Upsilon_0=\mathop{{\rm diag}}(-1/2,(1-\rme^{\rmi\omega_1T})^{-1},
(1-\rme^{-\rmi\omega_1T})^{-1},
(1-\rme^{(-\mu+\rmi\Omega)T})^{-1},
(1-\rme^{(-\mu-\rmi\Omega)T})^{-1}).
\]
Writing
$\vec\Theta(\tau)=\vec\Theta_0(\tau)+\ep\vec\Theta_1(\tau)+O(\ep^2)$,
we find that
\[
\vec\Theta_0(\tau)=\vec P_1(T)\calS(\tau)+
\frac{2(1-k)\vec Q_1(a_0(\tau)T)}{LC}
-\frac{(1+k-2ka_0(\tau))\vec Q_1(T)}{LC}
+\frac{2k}{TLC}\vec Q_2(T)
\]
and
\bean
\vec\Theta_1(\tau)&=&\frac{1}{T}\vec P_2(T)\calS'(\tau)
+\vec P_1(T)a_0(\tau)\calS'(\tau)
+T\vec P_0(T)a_0'(\tau)\calS(\tau)\\
&&{}+
\frac{2\left(k\vec Q_1(T)+(1-k)T\vec Q_0(a_0(\tau)T)\right)
(a_1(\tau)+a_0'(\tau))}{LC}
-\frac{(1+k-2ka_0(\tau))T\vec Q_0(T)a_0'(\tau)}{LC}.
\eean

The leading terms in \eqref{eq:ae} are those at $O(\ep^{-1})$, which give
\beq
\vec\gamma^TR\rme^{a_0(\tau)\Lambda T}
\rmD^{-1}
\left\{\Upsilon_{-1}\rme^{-a_0(\tau)\Lambda T}R^{-1}\vec\Theta_0(\tau)\right\}
=0.
\label{eq:ae-1}
\eeq
This equation may be considerably simplified by noting that 
$\Upsilon_{-1}\rme^{-a_0(\tau)\Lambda T}=\Upsilon_{-1}$ and, further, that
\beq
\Upsilon_{-1}\rme^{-a_0(\tau)\Lambda T}R^{-1}=\calR,
\label{eq:Up}
\eeq
where $\calR$ is a $5\times5$ matrix whose first row is $\vec v_1$ and
whose remaining elements are all zero. Thus we may satisfy \eqref{eq:ae-1}
by imposing the condition
\beq
\vec v_1\vec\Theta_0(\tau)=0.
\label{eq:forT0}
\eeq
It is readily established that
\[
\vec v_1\vec P_n(t)=-\frac{t^n}{n!LC},\qquad
\vec v_1\vec Q_n(t)=-\frac{t^n}{n!},
\]
and hence, from \eqref{eq:forT0},
\beq
a_0(\tau)={\textstyle\frac12}(1+\calS(\tau)),
\label{eq:a0}
\eeq
which is the analogue of the duty-cycle condition \eqref{eq:duty}.

The next terms to consider in \eqref{eq:ae} are those at $O(1)$. After
benefiting from the considerable simplification that follows from using
\eqref{eq:Up} and imposing \eqref{eq:forT0}, we find
\[
\vec\gamma^TR\rme^{a_0(\tau)\Lambda T}\left(
\Upsilon_0\rme^{-a_0(\tau)\Lambda T}R^{-1}\vec\Theta_0(\tau)+\rmD^{-1}
\calR\vec\Theta_1(\tau)\right)
=-1+2a_0(\tau).
\]
Then, since $\exp(\pm a_0(\tau)\Lambda T)$ and $\Upsilon_0$ are all
diagonal matrices, we see that
\[
\rme^{a_0(\tau)\Lambda T}\Upsilon_0\rme^{-a_0(\tau)\Lambda T}
=\Upsilon_0.
\]
Thus, by making use of this result and \eqref{eq:Up}, we have
\[
\vec\gamma^TR\left(
\Upsilon_0R^{-1}\vec\Theta_0(\tau)+\rmD^{-1}\calR\vec\Theta_1(\tau)\right)
=-1+2a_0(\tau),
\]
which we may solve by taking
\beq
\vec\gamma^TR\calR\vec\Theta_1(\tau)=
\calS'(\tau)-
\vec\gamma^TR\Upsilon_0R^{-1}\vec\Theta_0'(\tau),
\label{eq:forT1}
\eeq
where we have used \eqref{eq:a0} to eliminate $a_0(\tau)$.

Now $\calR\vec\Theta_1(\tau)=(\theta_1,0,0,0,0)^T$, where
\[
\theta_1=\vec v_1\vec\Theta_1(\tau)=
\frac{T}{LC}\left(g_1-\frac12(1-k)\calS(\tau)\calS'(\tau)\right),
\]
where $g_1$ is defined in \eqref{eq:g1def}.
Furthermore, elementary matrix algebra gives
\[
\vec\gamma^TR\calR\vec\Theta_1(\tau)=\theta_1\vec\gamma^T\vec w_1=
-\frac{LC}{\omega_1^2}(c_1\omega_1^2+c_3)\theta_1.
\]
The right-hand side of \eqref{eq:forT1} may be expressed more concretely by
noting that
\[
\vec\Theta_0'(\tau)=\vec P_1(T)\calS'(\tau)
+\frac{(1-k)T}{LC}\vec Q_0(a_0(\tau)T)\calS'(\tau)
+\frac{k}{LC}\vec Q_1(T)\calS'(\tau).
\]
If we now define
\[
p_n(t)=\vec\gamma^TR\Upsilon_0R^{-1}\vec P_n(t),\quad
q_n(t)=\vec\gamma^TR\Upsilon_0R^{-1}\vec Q_n(t),
\]
then $g_1$ is given by
\beq
g_1=\frac{(1-k)\calS(\tau)
\calS'(\tau)}{2} -
\frac{\displaystyle\omega_1^2
\left(1-\psi(\tau)\right)\calS'(\tau)}{(c_1\omega_1^2+c_3)T},
\label{eq:ga1is}
\eeq
where
\[
\psi(\tau)=
p_1(T)+\frac{(1-k)T}{LC}q_0({\textstyle\frac12}(1+\calS(\tau))T)
+\frac{k}{LC}q_1(T).
\]
This expression for $g_1$ enables us to determine the most significant
components of the audio distortion.

We note that without RC (i.e., for $k=0$) the expression for $g_1$ is
nonlinear in $\calS$, and hence the output contains harmonic distortion
at $O(\ep)$.  We also see that with RC ($k=1$) $g_1$ becomes the much simpler 
expression
\[
g_1=- \frac{\displaystyle\omega_1^2 
\left(1-p_1(T)-q_1(T)/(LC)\right)}{(c_1\omega_1^2+c_3)T} 
\calS'(\tau),
\]
which involves only terms that are linear in $\calS$ (the first {\em 
nonlinear} terms, involving quantities such as $((\calS')^2)'$, which 
involve three derivatives, will arise first at $O(\ep^3)$ in the 
output, cf.~\cite{CM15}).

Thus we have determined explicitly (at least, to $O(\ep)$) the way in 
which the nonlinear behaviour of the amplifier tracks the slowly varying 
audio input, and in particular the resulting low-frequency components of 
the output. While the leading-order tracking result, from 
\eqref{eq:gais} and \eqref{eq:a0}, that $g_a\sim\calS$, is well known 
and is easily understandable from the duty-cycle balance in 
\eqref{eq:duty}, a comprehensive calculation of the type above is 
necessary to obtain a complete perturbative calculation of corrections.

Our results confirm the conclusions of the small-signal model that RC 
(almost) completely linearises the output.

Although the perturbation calculation described in this section can, in 
principle, be taken to higher order in $\ep$, in practice the algebra 
required for this fifth-order system rapidly becomes unmanageable, even 
using computer algebra. Fortunately, the dominant contributions to the 
distortion seem to be captured by the terms to $O(\ep)$, for reasonable 
parameter values.

\section{Results}\label{sec:res}

\begin{table}
\begin{center}
\begin{tabular}{ll}\hline\noalign{\smallskip}
$R = 8\Omega$     & 
$c_1=1.3318\times10^5$/s           \\ 
\noalign{\smallskip}\hline\noalign{\smallskip}
$C = 0.5169\mu$F  & 
$c_2=1.3763\times 10^{10}$/s${}^2$ \\ 
\noalign{\smallskip}\hline\noalign{\smallskip}
$L = 10\mu$H      & 
$c_3=-1.0747\times10^{14}$/s${}^3$ \\ 
\noalign{\smallskip}\hline\noalign{\smallskip}
$T=1/384000$s     & 
$\omega_1=1.3195\times10^5$rad/s   \\ 
\noalign{\smallskip}\hline
\end{tabular}
\end{center}
\caption{Parameter values used in simulations, unless otherwise specified.}
\label{tab:1}
\end{table}

For our first set of simulations, we take the parameter values in 
Table~\ref{tab:1}. These give stable steady-state operation, according to 
the criteria in Section~\ref{sec:stab}. We carry out simulation of the 
amplifier in Matlab Simulink and compare results with the small-signal 
transfer function in \eqref{eq:tf} and with analytical predictions of the 
audio output from \eqref{eq:gais} and \eqref{eq:ga1is}.

\begin{table}
\begin{center}
\begin{tabular}{lll}\hline\noalign{\smallskip}
Frequency (kHz) & Analytical & Numerical \\
\noalign{\smallskip}\hline\noalign{\smallskip}
2 & $5.247\times10^{-5}$ & $5.258\times10^{-5}$ \\
\noalign{\smallskip}\hline\noalign{\smallskip}
3 & $2.23\times10^{-6}$ & $1.52\times10^{-6}$ \\
\noalign{\smallskip}\hline\noalign{\smallskip}
4 & $1.25\times10^{-5}$ & $1.38\times10^{-5}$ \\
\noalign{\smallskip}\hline
\end{tabular}
\caption{Absolute value of the Fourier components at various harmonics of
the input $1$kHz sine wave: analytical results from \eqref{eq:gais} and
\eqref{eq:ga1is}, and numerical results from simulation.}
\label{tab:2}
\end{center}
\end{table}

In the absence of RC ($k=0$), we examine a sine wave input 
\begin{equation}
u(t)=u_*\sin(2\pi f t),
\label{eq:sine}
\end{equation}
with $u_*=0.8$ and $f=1$kHz. For the analytical result in 
\eqref{eq:gais}, we keep terms at $O(1)$ and $O(\ep)$; the output 
Fourier component at the fundamental frequency is then predicted to be 
$-0.01356-0.4\rmi$, while simulation gives $-0.0166-0.3988\rmi$. The 
absolute values of the Fourier components for the second, third and 
fourth harmonics are given in Table~\ref{tab:2}. Given the small 
amplitude of harmonics and the small number of terms kept in the 
perturbation analysis, these results represent very good agreement 
between theoretical and numerical results.

In the presence of RC, we may compare simulation results both with the 
perturbation calculation above, and also with the predictions of the 
small-signal model. Taking terms at $O(1)$ and $O(\ep)$, \eqref{eq:gais} 
predicts that the output contains only the fundamental. Its prediction of 
the amplitude of the output fundamental agrees exactly with predictions of 
the small-signal model, when the latter is appropriately truncated. When 
$u(t)$ is as in \eqref{eq:sine}, with $u_*=0.8$ and $f=1$kHz, the output 
Fourier component at the fundamental frequency is analytically 
$-0.0135-0.3987\rmi$, and from simulation $-0.0166-0.3988\rmi$. With $f$ 
instead $2$kHz, the corresponding results are $-0.0263-0.3949\rmi$ and 
$-0.0327-0.3952\rmi$. At lower amplitude, with $u_*=0.5$ (and $f=1$kHz), 
the results are $-0.0084-0.2492\rmi$ and $-0.0104-0.2492\rmi$. Again the 
agreement between theoretical prediction and simulation is very good. The 
largest harmonic in the output is measured to be less than $10^{-5}$, 
which confirms the effectiveness of RC in eliminating higher harmonics 
from the output.

\subsection{Effects of instability}

\begin{figure}
\begin{center}
\epsfig{file=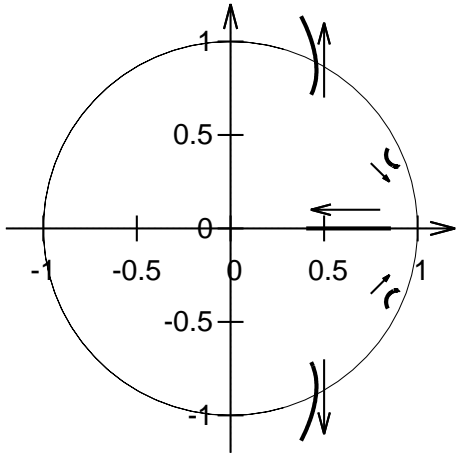,width=0.6\linewidth}
\end{center}
\caption{Path of the eigenvalues of $\calM$ in the complex plane as $c_1$
is varied from $10^{5}$/s to $4.5\times10^{5}$/s, all other parameter
values being as in Table~\ref{tab:1} and with $k=0$. Arrows indicate
the direction for increasing values of $c_1$.}
\label{fig:eigpat}
\end{figure}

The bifurcation structure of a negative-feedback pulse-modulated system 
such as described in this paper can be extremely 
intricate~\cite{amzg,azm} in response to a sinusoidal reference. 
However, the practical mode of operation for the present device aims to 
avoid instability; thus it is sufficient to use an approximation to the 
stability boundary, as we now describe. From either \eqref{eq:det} or 
\eqref{eq:detsca} we may determine whether the steady-state operating 
point in response to a \textit{constant} input $u_0$ is stable or 
unstable. We find (either with or without RC) that the stability 
boundary depends only very weakly on the value of $u_0$. 
Correspondingly, we find that the steady-state stability threshold gives 
a very good indication of the stability of operation in response to an 
audio sine-wave input (i.e., one that varies slowly compared with the 
time scale of the switching). The behaviour of the amplifier is markedly 
different in the ``stable'' and ``unstable'' cases, and in practice the 
threshold between the cases is quite sharp.

\begin{figure}
\begin{center}
\epsfig{file=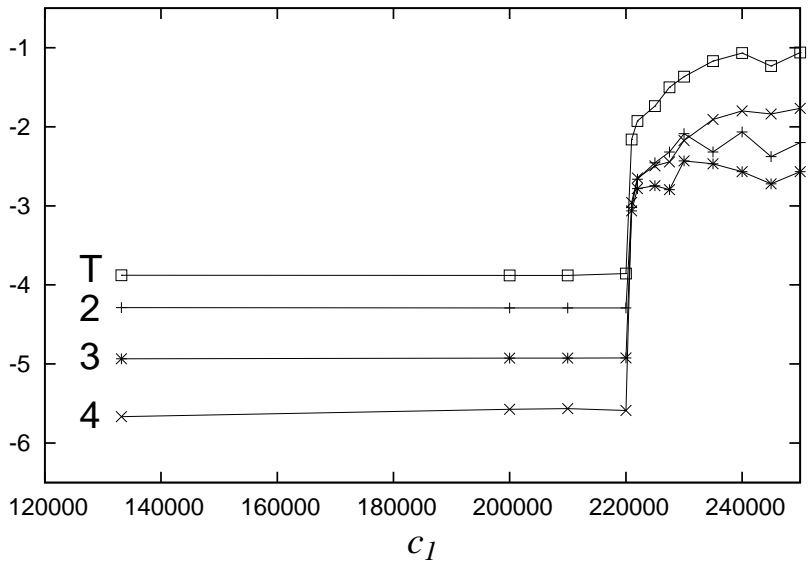,width=0.7\linewidth}
\end{center}
\caption{Line \textsf{T} shows $\log_{10}{\rm THD}$ as a function of $c_1$,
all other parameters being as in Table~\ref{tab:1}. Lines 
$2$, $3$, $4$, respectively, give $\log_{10}$ of the amplitudes of the
second, third and fourth harmonics in the output. The input is 
$u(t)=0.8\sin(2\pi f t)$, with $f=1$kHz.}
\label{fig:instab}
\end{figure}

For expository purposes, we use $c_1$ as our bifurcation parameter, 
holding all other parameters fixed at their values as in 
Table~\ref{tab:1}. The paths of the eigenvalues of the matrix $\calM$ 
are shown in Figure~\ref{fig:eigpat} as $c_1$ is varied from $10^{5}$/s 
to $4.5\times10^{5}$/s, for a constant input $u_0=0$, with $k=0$. In 
fact the eigenvalues vary little with the choice of $u_0$ or $k$. We 
find that the steady-state operating point is stable for $c_1<c_{1c}$, 
where $c_{1c}$ varies between $2.206\times10^5$/s and 
$2.208\times10^5$/s as $u_0$ varies in the interval $[-1,1]$. 
\textit{For practical purposes}, it is thus a reasonable approximation 
to consider that there is a single point at which the bifurcation from 
stability to instability takes place (although a more detailed analysis 
would undoubtedly reveal a rich, finer-grained bifurcation 
structure~\cite{amzg,azm}). Instability of the steady-state operating 
point arises through a pair of complex conjugate eigenvalues of 
$\mathcal M$ leaving the unit circle, as in Figure~\ref{fig:eigpat}. 
Beyond the bifurcation point, there are corresponding oscillations in 
the duty cycle, which grow until $a_n$ reaches 0 or 1, at which point 
the duty cycle saturates, and we observe one or more switching periods 
in which no switching in fact takes place (one or more pulses are 
``skipped''~\cite{BBCK08,dt02}). This saturation of the duty cycle tends 
to occur most readily when $|u(t)|$ is greatest. The oscillations in 
$a_n$ and its saturation at 0 or 1 lead to a sudden calamitous jump in 
the amplitude of harmonics in the output, and consequent sudden steep 
rise in the total harmonic distortion (THD); see 
Figure~\ref{fig:instab}. The THD of a signal may be defined as follows:
\[ 
\mbox{for\ }f(t)=\sum_{-\infty}^\infty f_n{\rme}^{n\rmi\omega 
t},\qquad \mbox{THD}=\frac{\sqrt{|f_2|^2+|f_3|^2+\cdots}}{|f_1|}. 
\]

As is evident in Figure~\ref{fig:instab}, there appears to be some 
uncertainty in our measurements of the harmonic amplitudes and the THD 
beyond the onset of instability, in contrast to our crisp results up to 
that point. The reason is that, prior to the onset of instability there 
are just two frequencies in the system, one associated with the audio 
sine wave and the other associated with the switching. Both frequencies 
are at our disposal; we choose these two frequencies to be commensurate 
and ensure that the time interval of simulation is an integer multiple 
of both the switching period $T$ and the sine-wave period $1/f$.  
However, beyond the bifurcation point a third frequency is present in 
the simulations, relating to the oscillations in the duty cycle about 
the (now unstable) steady-state response. This third frequency arises 
dynamically in the system and is not a parameter at our disposal. Hence 
our simulations in general do not contain an integer number of periods 
of this oscillation. Consequently, there is \textit{spectral 
leakage}~\cite{h78} (absent before the instability) and our measurements 
of various harmonic components are correspondingly contaminated. 
However, beyond the bifurcation point the THD performance of the 
amplifier is so poor that a precise measurement of the THD is 
unnecessary: post-instability the amplifier is all but useless for 
high-fidelity reproduction.

Of course our exploration of the high-dimensional parameter space of this 
amplifier is extremely limited: it is entirely possible that by choosing 
different parameter values and/or varying different parameters we might 
find a supercritical bifurcation leading to oscillations that saturate at 
small amplitude beyond the point of instability. In this case, any rise in 
THD is likely to be far less dramatic than that observed here.

\section{Conclusions}\label{sec:conc}

We have analysed the nonlinear response of a fifth-order pulse-modulated 
negative-feedback system to a slowly varying input. For the application 
at hand, we have demonstrated quantitatively the effectiveness of the 
ripple compensation technique in reducing audio distortion that arises 
from switching in the negative feedback loop. Our approach complements 
the usual focus on instability and bifurcation of piecewise-smooth 
systems~\cite{BBCK08}; our interest is in the accuracy with which the 
system tracks its input, which provides a challenging perturbation 
problem in its own right. We choose system parameters deliberately to be 
of physical relevance, avoiding instability. Our principal results have 
been a small-signal model, which linearises about a steady-state point 
of operation, and a nonlinear perturbation calculation that avoids such 
a linearisation. While the former is a standard piece of the engineer's 
toolkit, the latter, much more powerful, calculation is not.

We emphasise that the techniques described here, particularly the fully 
nonlinear calculation presented in a quite general formulation in 
Section~\ref{sec:nl}, are applicable to a wide variety of other 
nonlinear pulse-modulated systems~\cite{gc98}.
 
\section*{Acknowledgments}

This research did not receive any specific grant from funding agencies in
the public, commercial, or not-for-profit sectors.

\clearpage
\section*{References}
 

\end{document}